\documentclass[preprint,12pt]{elsarticle}
\usepackage{xcolor,colortbl}
\usepackage{multirow}
\usepackage{subcaption}
\usepackage{graphicx}
\usepackage{float}
\usepackage{adjustbox}

\definecolor{LF027}{rgb}{0.3542,0.0277,0.1119}
\definecolor{LF101}{rgb}{1.0360,0.8253,0.0600}
\definecolor{LF105}{rgb}{1.0177,0.3795,0}
\definecolor{LF116}{rgb}{0,0.3695,0.4123}
\definecolor{LF120}{rgb}{0,0.0242,0.3715}
\definecolor{LF126}{rgb}{0.3224,0.0174,0.3005}
\definecolor{LF129}{rgb}{0.9561,0.9372,0.9524}
\definecolor{LF135}{rgb}{0.8494,0.0294,0.1340}
\definecolor{LF139}{rgb}{0.0416,0.2644,0.0899}
\definecolor{LF181}{rgb}{0.0928,0.0809,0.1793}
\definecolor{LF182}{rgb}{0.6580,0.0129,0.1375}
\definecolor{LF332}{rgb}{0.6801,0.0140,0.2078}

\definecolor{ST_1_0deg_B}{rgb}{0.1750,0.1256,0.1178}
\definecolor{ST_2_0deg_B}{rgb}{0.1271,0.1215,0.1102}
\definecolor{ST_3_0deg_B}{rgb}{0.1291,0.1291,0.1291}
\definecolor{ST_4_0deg_B}{rgb}{0.1776,0.1292,0.1327}
\definecolor{ST_1_20deg_B}{rgb}{0.5950,0.5886,0.1783}
\definecolor{ST_2_20deg_B}{rgb}{0.1799,0.1651,0.1767}
\definecolor{ST_3_20deg_B}{rgb}{0.3419,0.1732,0.1761}
\definecolor{ST_4_20deg_B}{rgb}{0.6953,0.4791,0.0983}
\definecolor{ST_1_45deg_B}{rgb}{0.6572,0.8528,0.7039}
\definecolor{ST_2_45deg_B}{rgb}{0.6436,0.5335,0.5254}
\definecolor{ST_3_45deg_B}{rgb}{0.9004,0.6109,0.5256}
\definecolor{ST_4_45deg_B}{rgb}{0.8250,0.9523,0.6468}

\definecolor{ST_1_0deg_W}{rgb}{0.4254,0.3908,0.5243}
\definecolor{ST_2_0deg_W}{rgb}{0.5423,0.5098,0.5135}
\definecolor{ST_3_0deg_W}{rgb}{0.4814,0.5179,0.5144}
\definecolor{ST_4_0deg_W}{rgb}{0.3524,0.3871,0.4980}
\definecolor{ST_1_20deg_W}{rgb}{0.6740,0.6564,0.4838}
\definecolor{ST_2_20deg_W}{rgb}{0.4863,0.4673,0.4785}
\definecolor{ST_3_20deg_W}{rgb}{0.4990,0.4702,0.4660}
\definecolor{ST_4_20deg_W}{rgb}{0.7442,0.5709,0.4246}
\definecolor{ST_1_45deg_W}{rgb}{0.8334,0.9562,0.8990}
\definecolor{ST_2_45deg_W}{rgb}{0.7930,0.6903,0.6824}
\definecolor{ST_3_45deg_W}{rgb}{0.8967,0.6819,0.6312}
\definecolor{ST_4_45deg_W}{rgb}{0.8574,0.9689,0.7438}

\usepackage{amssymb}
\usepackage{draftwatermark}

\journal{Solar Energy}

\begin{document}

\SetWatermarkColor[gray]{0.6}
\SetWatermarkText{Preprint submitted to $Solar Energy$}
\SetWatermarkAngle{90}
\SetWatermarkHorCenter{18.5cm}
\SetWatermarkFontSize{0.5cm}

\begin{frontmatter}



\title{Efficient color coatings for single junction and multijunction colored solar cells}

\author[1]{Farid Elsehrawy\corref{cor1}}
\author[2]{Konrad Klockars}
\author[2]{Orlando J. Rojas}
\author[1]{Janne Halme}

\affiliation[1]{organization={Department of Applied Physics, Aalto University},%Department and Organization
            addressline={Puumiehenkuja 2}, 
            city={Espoo},
            postcode={02150}, 
            country={Finland}}

\affiliation[2]{organization={Department of Bioproducts and Biosystems, Aalto University},%Department and Organization
        	addressline={Otakaari 1}, 
        	city={Espoo},
        	postcode={02150}, 
        	country={Finland}}

\cortext[cor1]{Corresponding author: Farid Elsehrawy, $farid.elsehrawy@aalto.fi$}

\begin{abstract}
Colored solar cells suffer from lower efficiency due to reflection and absorption losses by coatings. By studying different types of coatings on solar cells, the spectrum parameters impacting the solar cell efficiency were identified. A collection of color coatings was fabricated and characterized using spectrophotometry and colorimetric photography. The coatings include commercial absorption filters, commercial distributed bragg reflectors, lab-made interference coatings, and cellulose nanocrystal coatings. Commercial single junction and multijunction solar cells were used to measure the impact of the color coatings on solar cell performance. Structural colors were found to result in the highest brightness to color loss ratio. Structural colors do not fade provided that the structure is not compromised. Color loss in single junction solar cells is a result of the proportion of reflected light in the absorption range of the material. Multijunction solar cells were strongly affected by current mismatch losses, where narrow reflection peaks resulted in high efficiency losses. Structural color coatings made using cellulose nanocrystals exhibited low and broad reflection peaks that produced bright colors and maintained high performance of single junction and multijunction solar cells, retaining up to 83 \% of the reference power obtained in the absence of a color coating.
\end{abstract}


% \begin{highlights}
% \end{highlights}


\end{frontmatter}


\section{Introduction}
\label{sec:intro}

In the past decades, the number of global photovoltaic system installations has been exponentially increasing to meet the demand for sustainable energy sources. The majority of photovoltaic system installations are in the utility segment that comprises solar farms, with a share of added solar PV capacity standing at 69 \% in 2021 \cite{renewables2020technical}. Meanwhile, the share of residential and commercial/industrial PV capacity additions remain much lower at 13 \% and 17 \%, respectively. PV systems are predominantly deployed in solar farms and rooftops due to the available space and ease of installation, with 60 \% of the currently installed PV capacity being in power plant systems and 35 \% in rooftop installations \cite{equipment2018international}. However, in urban areas where land and rooftop space are limited, building-integrated photovoltaics (BIPV) are promising alternatives due to their utilization of space on the building envelope (i.e., windows and façades).

There are still several barriers to the widespread adoption of BIPV, including their low efficiency, high cost of BIPV modules, and need for customized solutions \cite{yang2016building}. Moreover, BIPV are preferred to be visually appealing by matching the colors of their surrounding environment \cite{bao2017understanding}. To address the visual appeal of BIPV, colored photovoltaics can be used to provide aesthetic façades that also enable energy harvesting. 

Various methods can be used to form color coatings for photovoltaics, such as using pigments \cite{eastaugh2007pigment}, structural \cite{chung2012flexible}, and plasmonic \cite{kristensen2016plasmonic} colors. The spectral absorption of a color coating varies depending on the reflected color \cite{peharz2018quantifying}, color lightness \cite{halme2019theoretical}, and coloration method. Pigment-based colors partly absorb visible (VIS) light and reflect or transmit the remaining light spectrum. Alternatively, structural colors originate from the selective reflection of light due to the physical periodic nanostructure (which can be formed by the assembly of nanoparticles). As structural colors do not contain pigments that absorb visible light, they reflect a small portion of the VIS spectrum and transmit the remainder, making structural coloration a promising path for colored photovoltaics. In addition to their superior optical performance, their colors do not fade over time provided that the structure is not compromised.

To show a certain color, a portion of the solar spectrum is reflected by the color coating. The color chromaticity depends on the reflected wavelengths and the color brightness depends on the amount of reflected VIS light. To meet the performance and aesthetics requirements for BIPV, it is necessary to develop efficient, scalable color coatings that can produce a sufficiently wide range of colors with varying brightness and chroma. Since the reflected energy is lost, color coatings can be detrimental to the solar cell efficiency. The loss due to a color coating, or color loss, depends on the amount of light reflected and absorbed by the coating and the spectral response of the used solar cell. 

The prevalent solar cells in the current BIPV market are based on crystalline silicon, which absorb photons with energy exceeding that of the silicon bandgap (1.12 eV). The absorption range of silicon is limited as it mostly encompasses the VIS and parts of the near-infrared (NIR) light spectrum, causing color coatings to have a strong impact on the solar cell efficiency. Alternatively, tandem solar cells can be promising due to their wider absorption range. Although tandem solar cells are commonly perceived to have high production costs, novel tandem configurations such as silicon-Perovskite solar cells are already in the process of achieving commercialization and low production costs \cite{sofia2020roadmap}.

\begin{figure}[h!]
    \centering
    \includegraphics[width=0.7\columnwidth]{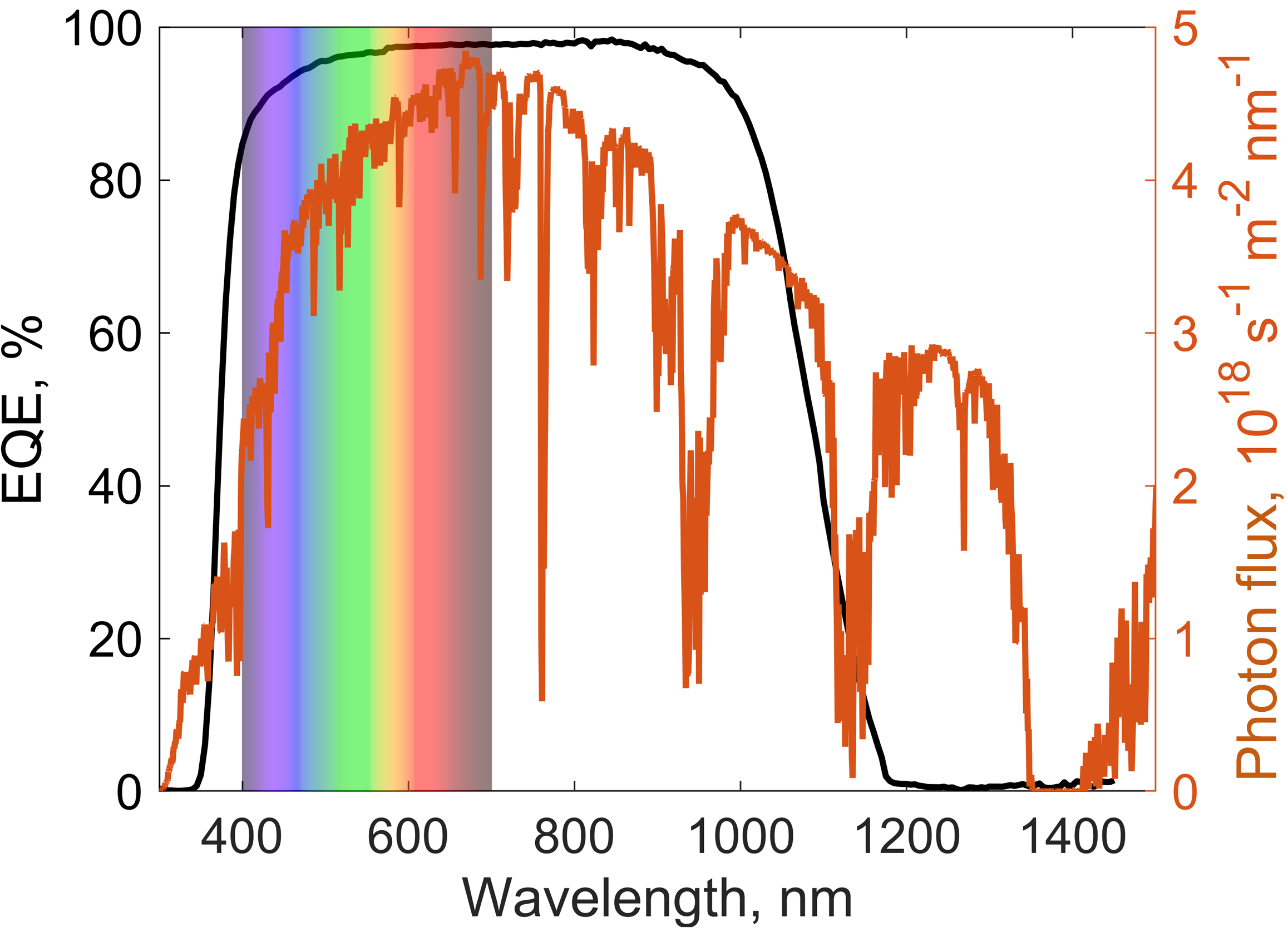}
    \caption{Left axis: measured spectral response (external quantum efficiency, EQE) of a commercial monocrystalline silicon solar cell (IXYS KXOB25-14X1F). Right axis: the spectral photon flux of AM1.5G illumination.}
    \label{fig:EQE_SPF_Spectrum}
\end{figure}

In this study, a commercial single junction (monocrystalline silicon) solar cell and a tandem solar cell were studied in combination with different color coatings. The spectral response of the commercial monocrystalline silicon solar cell is shown in Fig.\,\ref{fig:EQE_SPF_Spectrum} (left axis) and the spectral solar irradiance at air mass 1.5G (right axis), where the highest solar cell absorption and irradiated power occur around the VIS region (i.e. wavelengths ranging from 380 - 780 nm).

An ideal color coating would reflect the spectrum necessary to produce the desired color while transmitting the remaining light to be absorbed by the solar cell. The reflected light spectrum can take any shape, resulting in countless approaches to generate a given color. In the ideal case, the color would be generated by reflecting one or two narrow bands of visible light only at the wavelengths where the human eye is most sensitive in producing color sensation \cite{halme2019theoretical}. The ideal coating would have no parasitic absorption, thereby resulting in maximum light absorption by the solar cell. Commercially available color coatings can be relatively efficient by transmitting up to 84 \% of light while reflecting the desired colors \cite{jolissaint2017colored}. They are available in a wide range of colors \cite{KromatixColoredPV,SolaxessColoredPV,RacellColoredPV,DanishSolarEnergy}, allowing high design flexibility for architects. However, the color brightness comes at the expense of the efficiency: brighter reflected colors result in less efficient coatings \cite{halme2019theoretical,royset2020coloured}. This presents the main challenge of creating aesthetic colors for solar cells while maintaining high efficiency. 

It is crucial to study the characteristics that make color coatings efficient and aesthetic for different types of solar cells. An extensive study comprising color coatings that generate color with multiple approaches would enable identifying the key features of color coatings. In addition, using the same solar modules to characterize the performance of diverse color coatings allows comparing the studied coloration methods. The best coloration method should enable tuning of the transmitted spectrum to maximize the efficiency of different solar cell types.

In this study we investigated the performance of various color coatings for use in colored single junction (SJ) and multijunction (MJ) solar cells. The study included a systematic comparison of the three characteristically different ways to produce color: absorption filters, interference filters, and photonic crystals. The color coatings were optically characterized using spectrophotometry and their perceived colors were characterized using colorimetric photography. Of the studied samples, the biobased photonic crystal colors were found to be the most promising in terms of optical efficiency, tunability, and scalability of production.

In addition to optically characterizing color coatings, the implications of using color coatings with single junction and tandem solar cells were investigated. Measuring the performance of different solar cells with and without the color coatings allowed quantifying the impact of color coatings on each solar cell. The performance of the colored solar cells was experimentally characterized by covering the solar cells with optical coatings fabricated using different methods and measuring the electrical parameters of the cells. The optical coatings include pigment-based absorption filters, dielectric multilayer interference filters, and self-assembling photonic crystals. Two commercial solar cells were used: a SJ monocrystalline silicon solar cell and a tandem GaInP$_2$/InGaAs/Ge heterojunction solar cell. The single junction solar cell (SJSC) was found to exhibit proportionally lower performance depending on the transmittance of the color coating. However, the multijunction solar cell (MJSC) exhibited significantly lower performance due to current mismatch losses caused by the color coating. Nevertheless, MJSCs retained higher efficiencies when used with structural color coatings.

\section{Materials}
\label{sec:materials}

The color coating samples contained multiple coating types to provide a wide-ranging study. First, absorption-based optical filters in several colors (see Table \ref{tab:LF_Colors}) were used to model the effect of various transmittance spectra on the solar cell performance. The absorption-based optical filters are created using pigments, and were obtained from a commercial source (Lee Filters). Despite acting as light absorption filters (for lighting applications), they are useful to simulate optical coatings with a wide range of transmittance spectra.

\begin{table}[h]
    \centering
    \begin{tabular}{|l|c|c|c|c|}
    \hline
      \multirow{2}{*}{Sample}   &  \multicolumn{4}{c}{Color coordinates} \\     \cline{2-5}
       & $L*$ & $a*$ & $b*$ & Color  \\
      \hline
     LF 027, Medium Red & 17.6 & 36.6 & 11.5 & \cellcolor{LF027} \\
     LF 101, Yellow & 86.7 & 4.4 & 86.0 & \cellcolor{LF101}\\
        LF 105, Orange&62.2&58.8&71.9 & \cellcolor{LF105}\\
        LF 116, Medium Blue-Green&35.8&-21.3&-13.5 & \cellcolor{LF116}\\
        LF 120, Deep Blue&7.1&33.5&-52.8 & \cellcolor{LF120}\\
        LF 126, Mauve&17.6&40.9&-22.8 & \cellcolor{LF126}\\
        LF 129, Heavy Frost&94.9&2.3&-1.3 & \cellcolor{LF129}\\
        LF 135, Deep Golden Amber&45.6&70.6&46.5 & \cellcolor{LF135}\\
        LF 139, Primary Green&24.3&-28.7&21.6 & \cellcolor{LF139}\\
        LF 181, Congo Blue&8.2&9.4&-16.4 & \cellcolor{LF181}\\
        LF 182, Light Red&35.0&58.8&32.1 & \cellcolor{LF182}\\
        LF 332, Special Rose Pink&36.5&60.9&22.9 & \cellcolor{LF332}\\
    \hline
    \end{tabular}
    \caption{Names and $L^*a^*b^*$ color coordinates of the absorption-based optical filters used in the study. The color coordinates are obtained by color-corrected photography against a diffuse white background.}
    \label{tab:LF_Colors}
\end{table}

Next, the study included various implementations of structural colors. Structural colors allow forming the desired color by using surfaces with layered or periodic structures where the layer thickness or structural periodicity is close to the wavelength of visible light, thereby causing interference. Various photonic structures can be used to form structural colors, including dielectric multilayers \cite{boudaden2005multilayered}, diffraction gratings \cite{uddin2012efficient}, and bi-periodic photonic crystals \cite{cho2009two}. The first structural color approach investigated in this work was to periodically vary the refractive index of the material by depositing dielectric multilayer films to form interference coatings. Such coatings allow reflecting light in one or multiple selected wavelengh bands to produce the desired color. The main challenge with deposition of dielectric multilayer films is the process complexity, which generally involves specialized deposition equipment. 

\begin{figure}[h!]
    \centering
    \includegraphics[width=0.5\columnwidth]{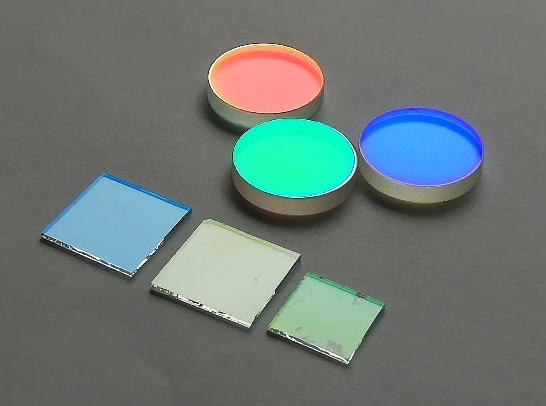}
    \caption{Red, green, and blue commercial reflective color filters (cylindrical) obtained from Edmund Optics (EO) and the blue, yellow-green, and green structural color samples (rectangular) fabricated using electron beam deposition of SiO$_2$ – TiO$_2$ multilayer stacks.}
    \label{fig:ST_Samples}
\end{figure}

The interference coating samples included a set of reflective color filters that were obtained from a commercial source, Edmund Optics (EO). The reflective color filters were designed to reflect the colors red, green, and blue, respectively, in the specular direction while transmitting the remaining VIS wavelengths. The interference coating samples also included a set of lab-made color filters, fabricated by electron beam deposition of SiO$_2$ and TiO$_2$ multilayers on fused silica substrates. Three samples were included in the study, shown in Fig.\,\ref{fig:ST_Samples}, comprising a 2-layer blue filter, 4-layer yellow-green filter, and a 6-layer green filter. The samples were formed by depositing layers of varying thicknesses; the blue filter was formed by 20 nm TiO$_2$ and 167 nm SiO$_2$, the yellow-green filter was formed by 15 nm TiO$_2$, 140 nm SiO$_2$, 10 nm TiO$_2$, and 70 nm SiO$_2$, and the green filter was formed by 8 nm TiO$_2$, 166 nm SiO$_2$, 6 nm TiO$_2$, 162 nm SiO$_2$, 4 nm TiO$_2$, and 65 nm SiO$_2$. SiO$_2$ was deposited at a rate of 1 Å/s and TiO$_2$ at 0.05 Å/s, both under a constant vacuum pressure of 1e-6 Torr. The layer thicknesses were optimized using an in-house developed script that performs electromagnetic simulations using the transfer matrix method and matches the reflectance spectrum to that of the target color.

Another investigated structural color approach involved cellulose nanocrystals (CNCs) that form colored iridescent films. The CNCs are rod-shaped nanoparticles chemically extracted from cellulosic fibres \cite{habibi2010cellulose};  in the present case they were extracted through hydrolysis using concentrated sulfuric acid. In aqueous dispersions, the CNCs are able to self-assemble into a liquid crystalline helical structure that can be preserved upon solvent evaporation \cite{schutz2020equilibrium}. The distance between the periodic 360$^\circ$ twists in the helical structure constitutes the pitch. The helical architecture interferes with light through Bragg reflection, resulting in angle-dependent visible color when the pitch approaches the wavelength of VIS light (380 – 780 nm)\cite{de1951rotatory}. The edges of the films or coated areas exhibit different colours due to the coffee-ring effect \cite{mu2015droplets}, arising from an elevated drying rate towards the sides and corners \cite{klockars2019asymmetrical}.

\section{Methods}
\label{sec:methods}

The samples were optically characterized using multiple methods. Spectrophotometry was used to obtain the total transmittance and total reflectance of the samples. The spectra were measured using the Agilent CARY 5000 UV-Vis-NIR spectrophotometer with the Diffuse Reflectance Accessory (DRA). It performs monochromatic measurements using two photodetectors (UV-Vis and NIR) to produce the transmittance or reflectance curve. The total transmittance was measured by placing the sample in front of the DRA and measuring the light transmitted through the sample into the integrating sphere. The total reflectance was measured by placing the sample at the back of the DRA with a light trap behind it.   

\begin{figure}[h!]
    \centering
    \includegraphics[width=0.4\columnwidth]{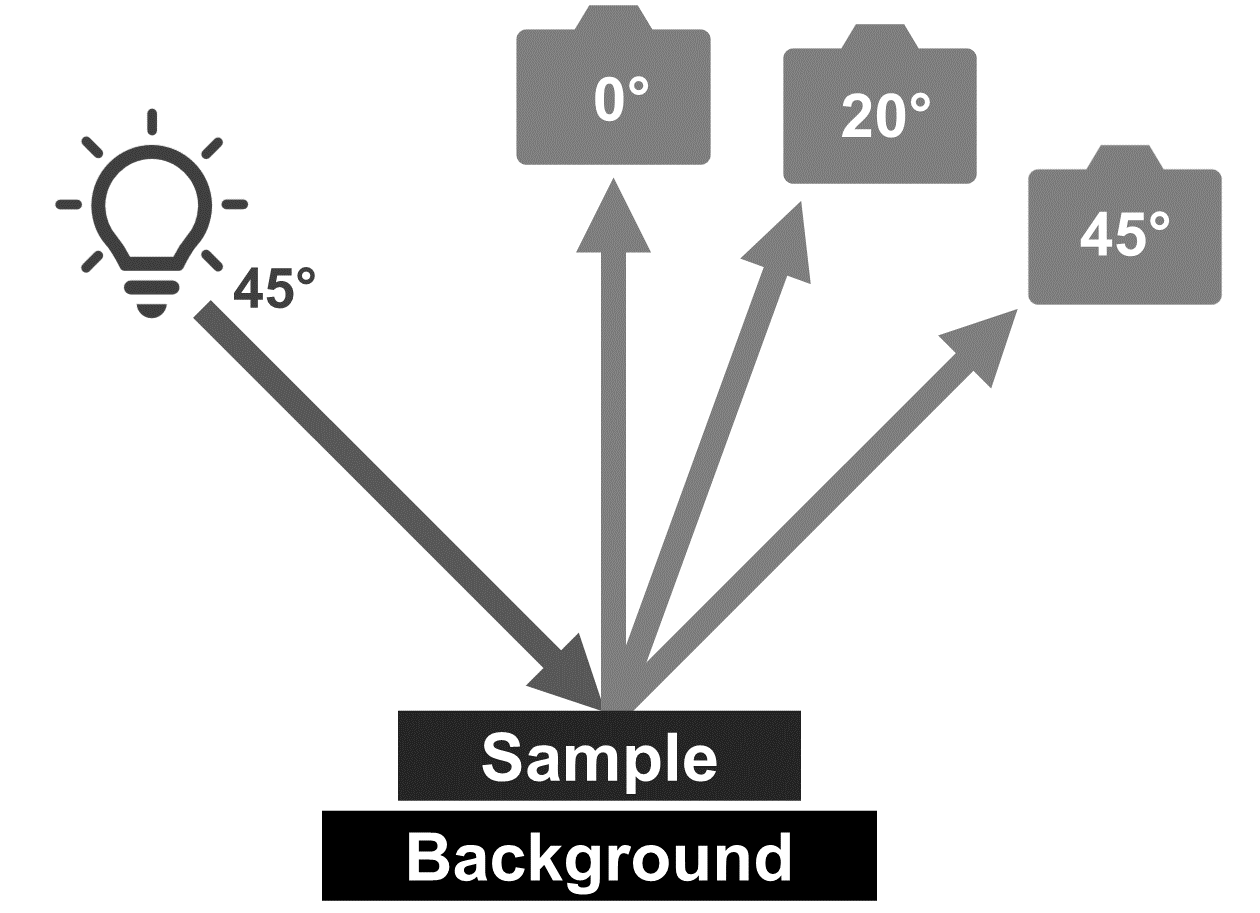}
    \caption{Diagram of the photography setup at different angles.}
    \label{fig:photography_setup}
\end{figure}

Next, color photography was used to estimate the color of the iridescent CNC samples. The photography was performed under controlled illumination conditions at varying angles, as shown in Fig.\,\ref{fig:photography_setup}, followed by color correction and color coordinate extraction.

\begin{figure}[h!]
    \centering
    \includegraphics[width=0.5\columnwidth]{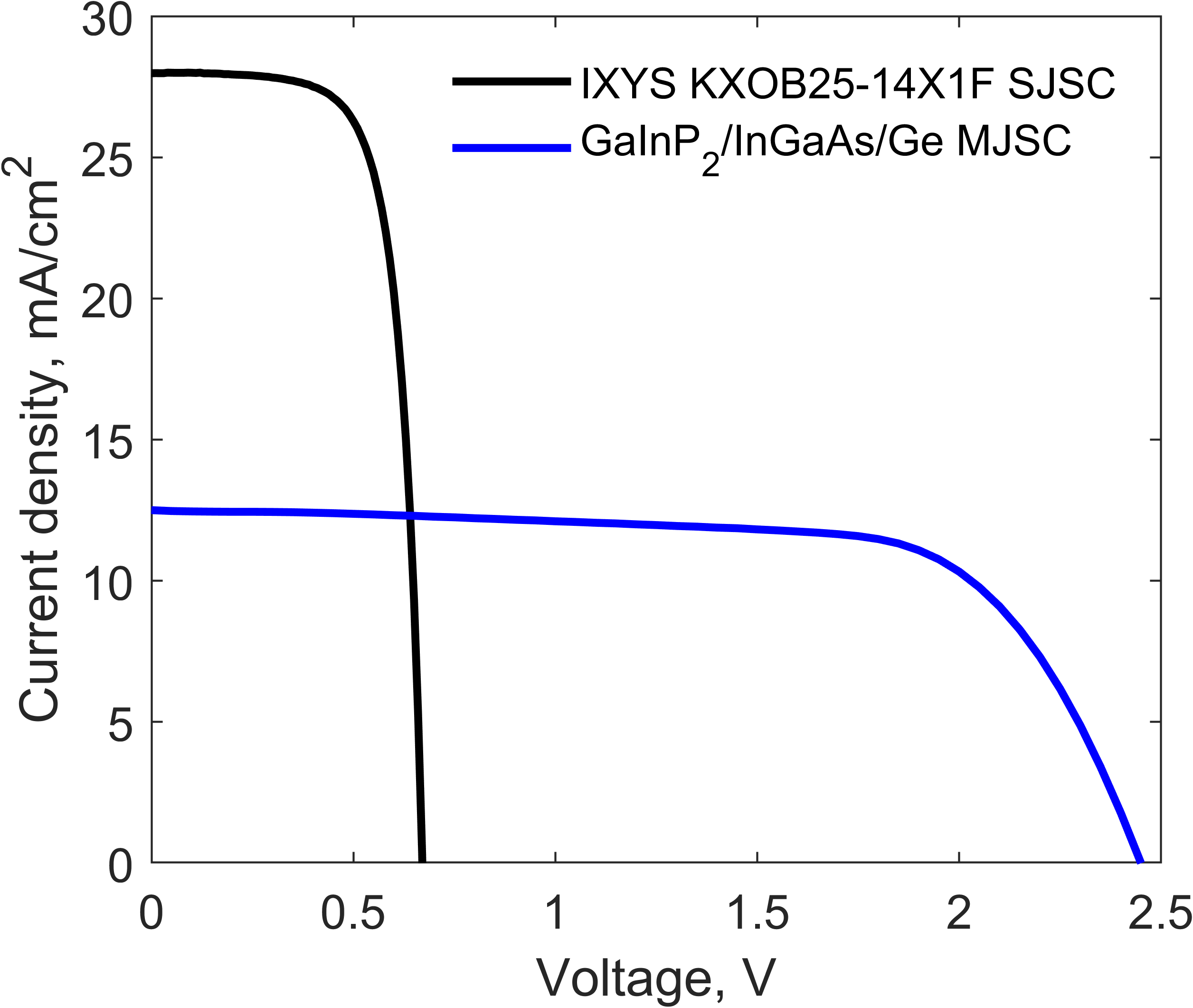}
    \caption{The current-voltage curve of the SJSC and MJSC used in the study.}
    \label{fig:IVs_bare_solar_cells}
\end{figure}

In addition to the optical characterization methods, electrical characterization was performed on the solar cells covered by the optical color coatings. The current-voltage characteristics were used to measure the performance of the solar cells when combined with the different color coatings. The measurements were performed using the Peccell solar simulator at 1 sun illumination (AM1.5G spectrum, 1000 W/m$^2$ irradiance). Two commercial solar cells were used: a monocrystalline SJSC (IXYS KXOB25-14X1F) and a GaInP$_2$/InGaAs/Ge MJSC (Fullsuns FR-G-CC-0505). The solar cells are not optimized for colored solar cells but were used as examples of commercially available solar cells. The IV curves of the bare reference solar cells are shown in Fig.\,\ref{fig:IVs_bare_solar_cells}. 

\section{Results}
\label{sec:results}

The various optical filters were first characterized using spectrophotometry to compare their optical performance. Next, their color was evaluated with colorimetric photography. Later, the solar cells covered with the optical filters were electrically characterized to evaluate the effect of the color coatings on the solar cell performance.

\subsection{Optical characterization}

\begin{figure}[h!]
    \centering
    \includegraphics[width=0.6\columnwidth]{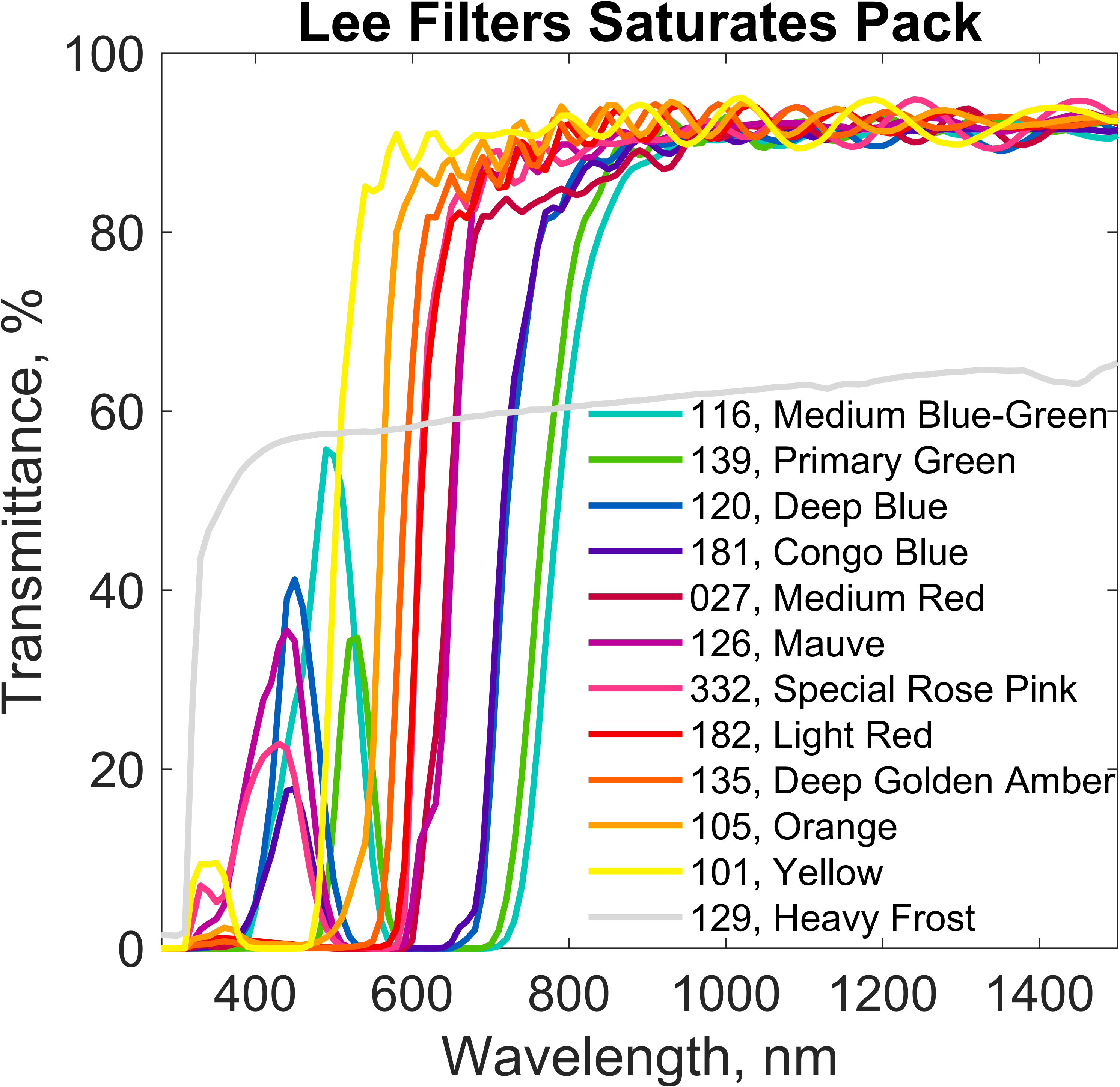}
    \caption{Measured total transmittance spectra of commercial absorption color filters from Lee Filters (Saturates Pack).}
    \label{fig:LF_transmittance}
\end{figure}

The transmittance spectra of the absorption filters, shown in Fig.\,\ref{fig:LF_transmittance}, demonstrates that the filters transmit light in the NIR range, which can be absorbed by the solar cell. Within the VIS range, the coatings transmit part of the spectrum depending on the color, which affects the solar cell performance differently. As for ultraviolet (UV) light, the transmittance is almost zero, indicating significant UV absorption in the pigments. Lighter colors tend to show higher transmittance in the VIS range due to the reduced light absorption by the pigments.

\begin{figure}[h!]
    \centering
    \includegraphics[width=0.45\columnwidth]{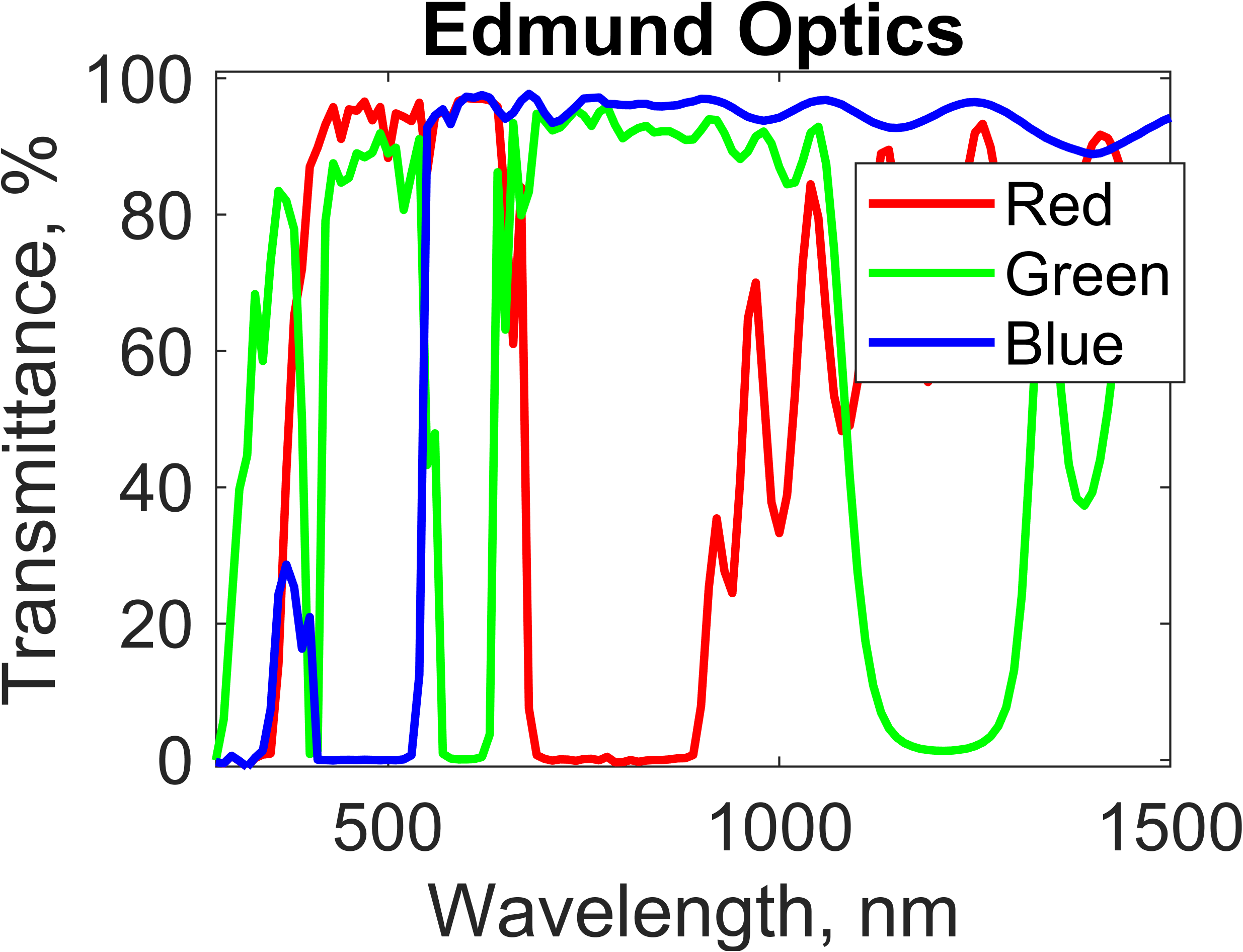}
    \includegraphics[width=0.45\columnwidth]{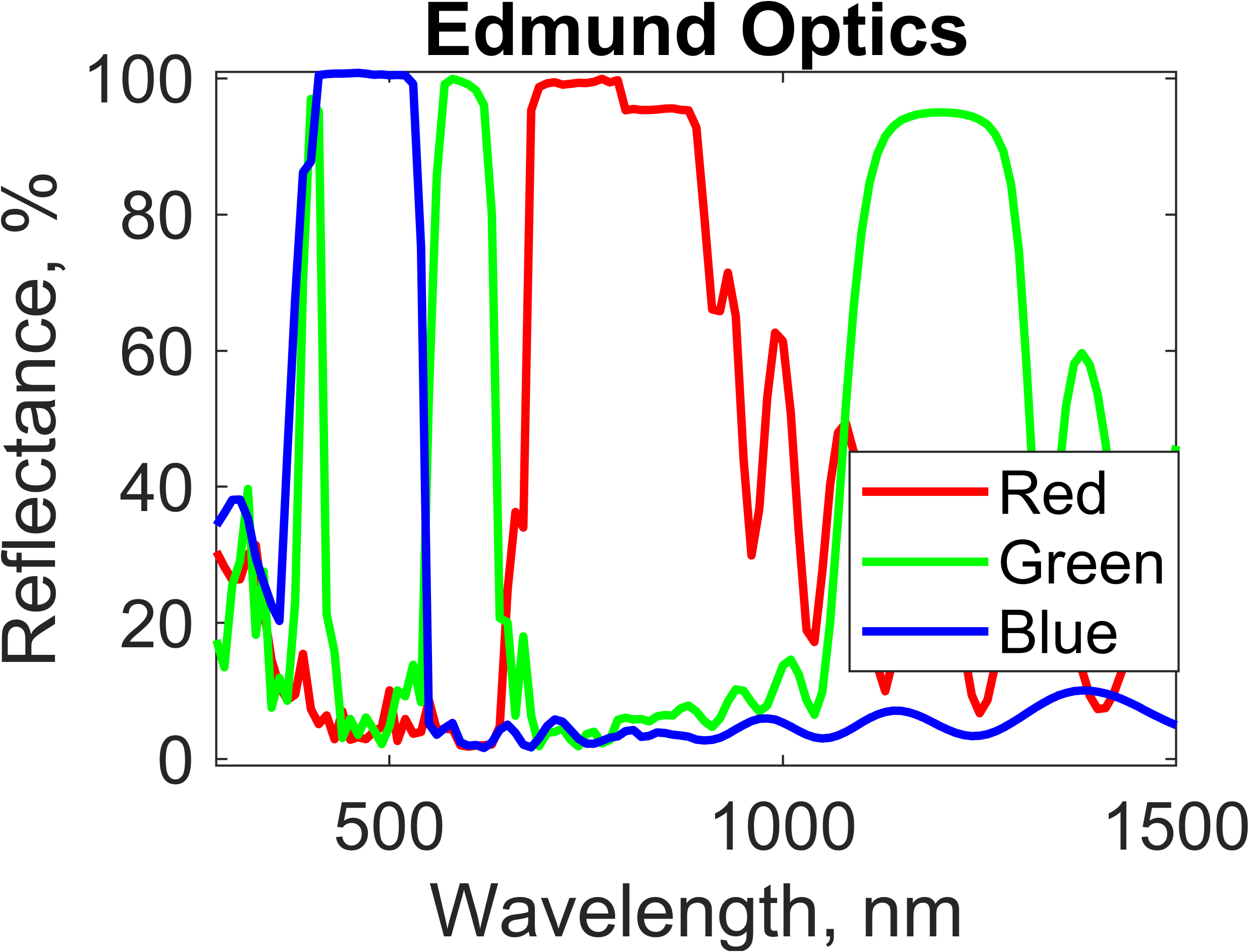}
    \caption{Measured total transmittance and total reflectance spectra of commercial reflective color filters (red, green, and blue) from Edmund Optics.}
    \label{fig:EO_spectra}
\end{figure}

The measured transmittance and reflectance spectra of the commercial interference coating samples are shown in Fig.\,\ref{fig:EO_spectra}. The samples demonstrated high reflectance (approximately 100 \%) at the wavelengths corresponding to the reflected color, and sufficiently high transmittance for other wavelengths. The green sample has two dips in the transmittance spectrum, corresponding to the reflected peaks that generate the green color. The dips were sufficiently narrow to allow light transmission to the solar cell. The blue sample formed the color by reflecting a single, moderately-wide peak centered around 450 nm. The red sample, however, reflected an exceptionally wide peak centered at 800 nm that causes high losses when compared to other colors. Both the red and green samples exhibited reflectance in the NIR range, significantly impacting the performance of solar cells absorbing NIR light.

\begin{figure}[h!]
    \centering
    \includegraphics[width=0.45\columnwidth]{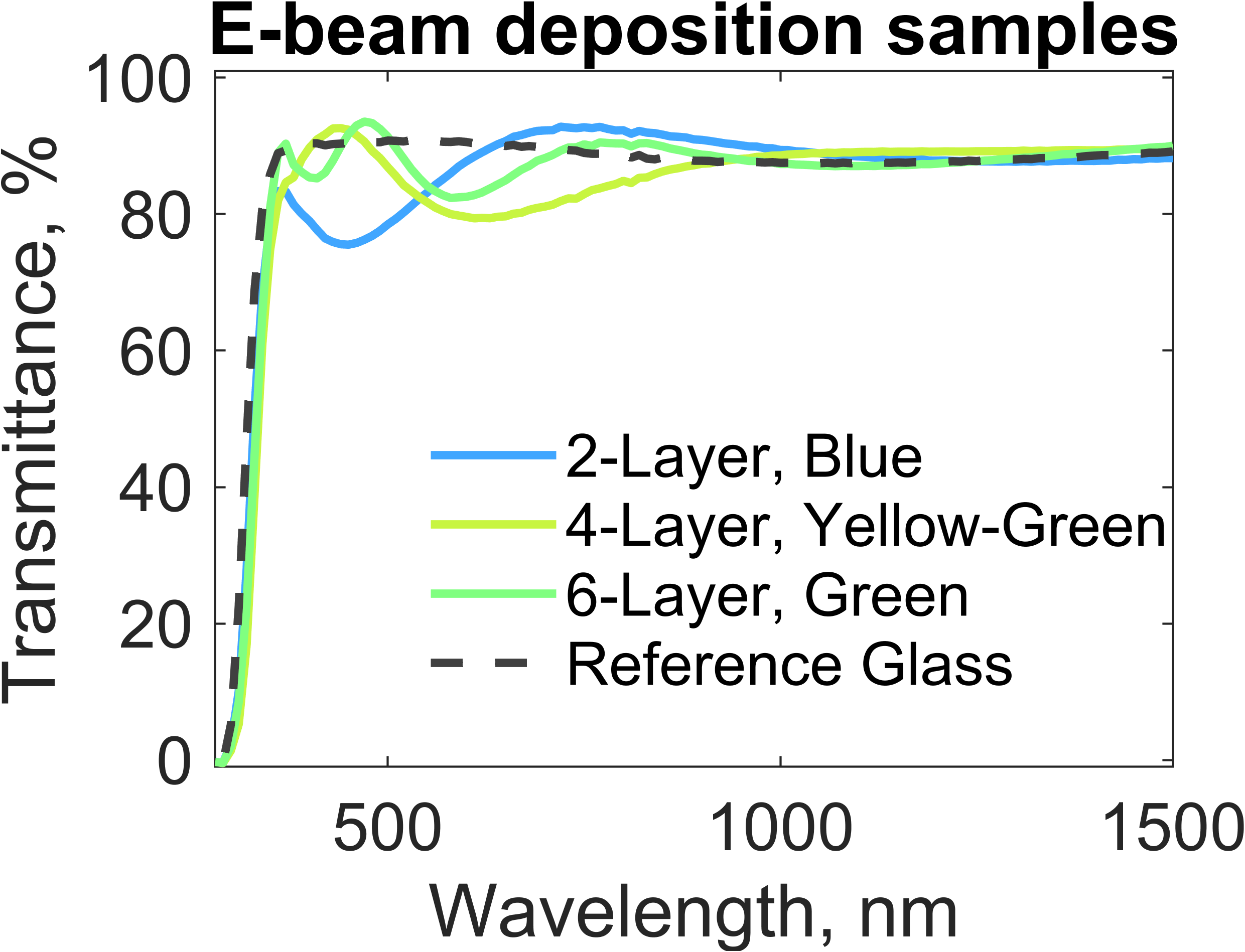}
    \includegraphics[width=0.45\columnwidth]{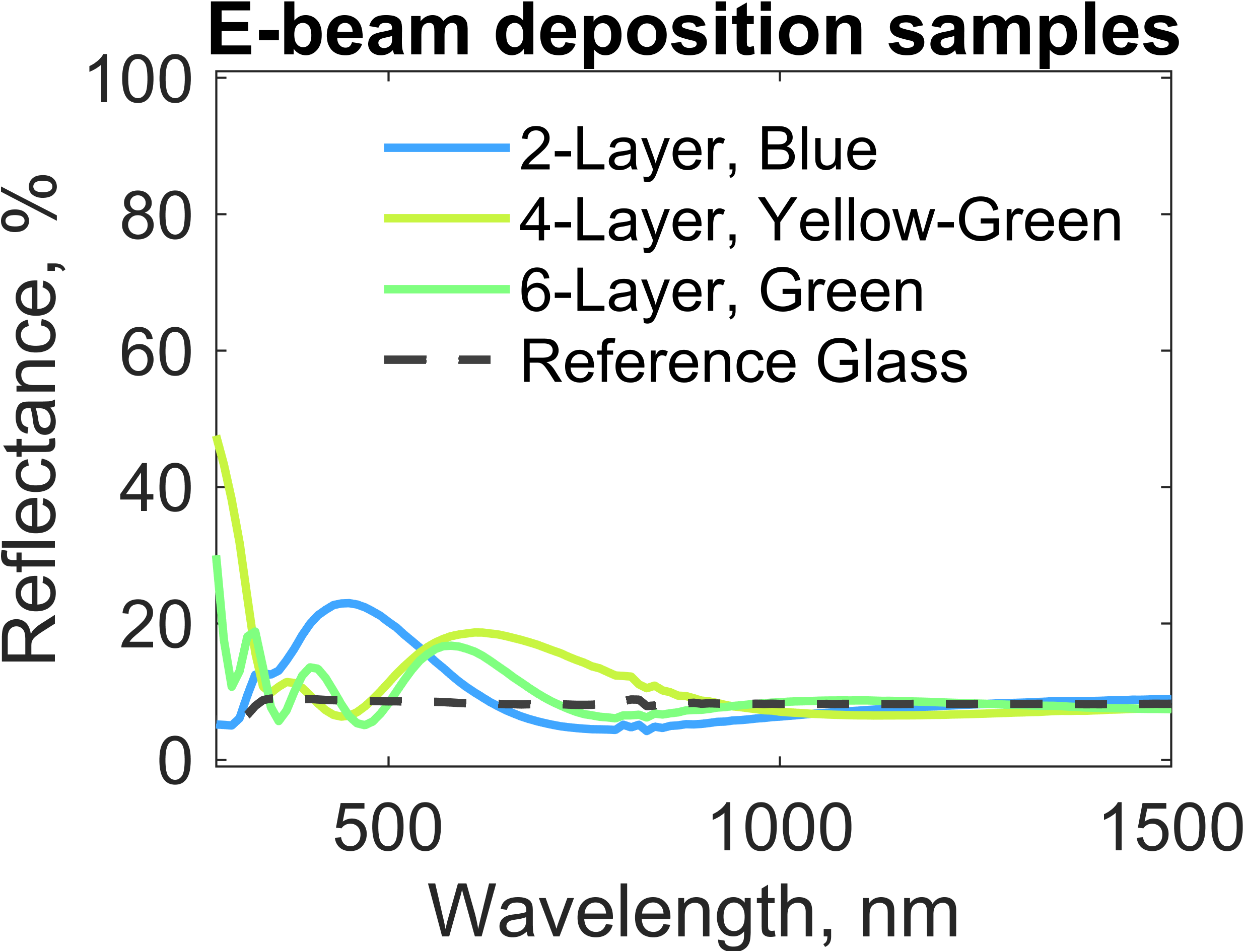}
    \caption{Measured total transmittance and total reflectance spectra of interference color filters fabricated using electron beam deposition of SiO$_2$-TiO$_2$ multilayers.}
    \label{fig:Ebeam_spectra}
\end{figure}

The next samples in the study were interference filters fabricated using electron beam (E-beam) deposition of SiO$_2$-TiO$_2$ multilayers, whose spectra are displayed in Fig.\,\ref{fig:Ebeam_spectra}. These filters generated color by reflecting light around a center wavelength corresponding to the desired color: 440 nm for the blue sample, 610 nm for the yellow-green sample, and 570 nm for the green sample. The filters were designed to reflect broad peaks, which produced bright colors while transmitting over 90 \% of the remaining spectrum. By examining the spectra of the silica substrate (Reference Glass sample), it can be observed that the substrate reflectance is around 10 \% and that the E-beam deposition coating is almost fully transparent outside the VIS region. 

\begin{figure}[h!]
    \centering
    \includegraphics[width=0.45\columnwidth]{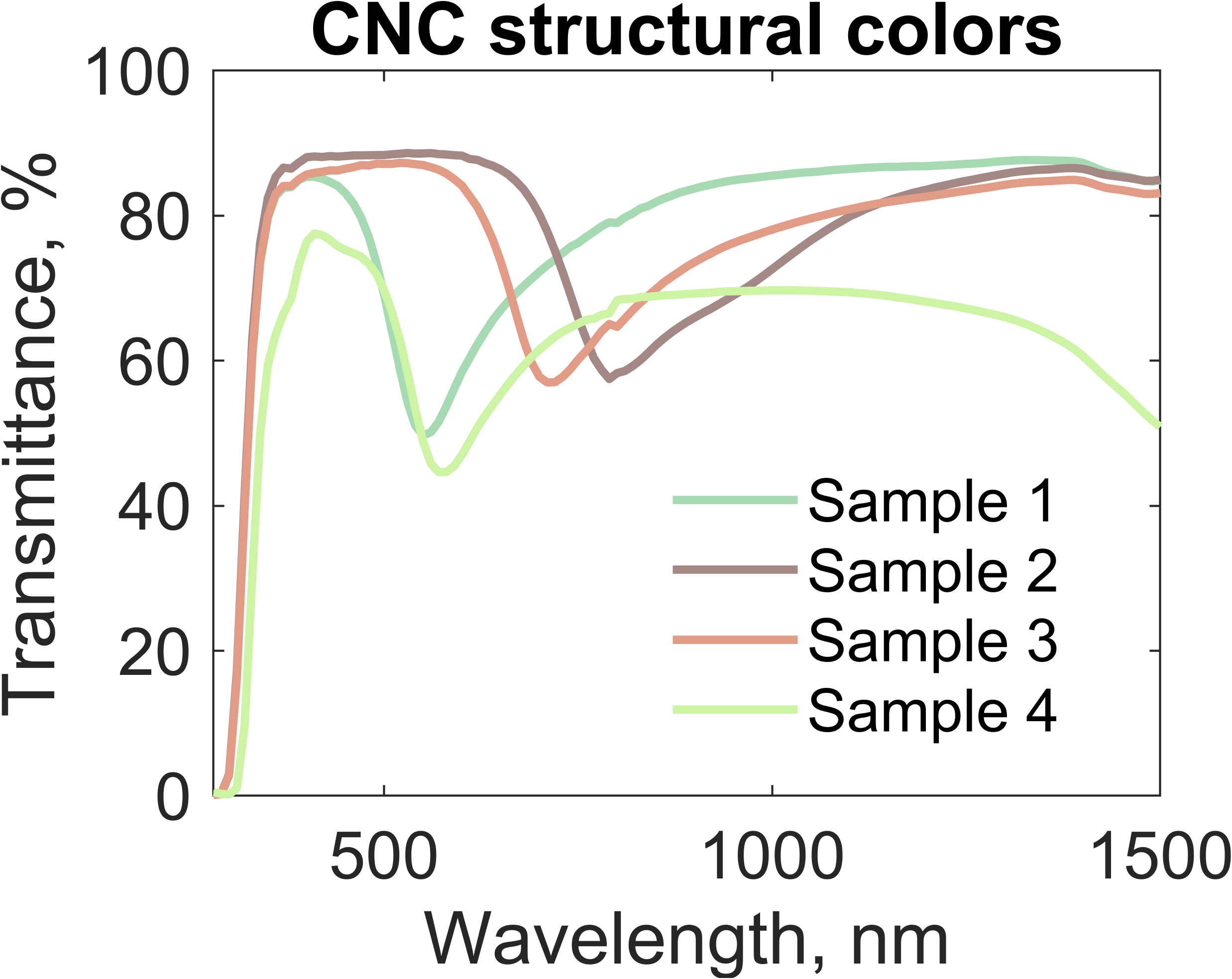}
    \includegraphics[width=0.45\columnwidth]{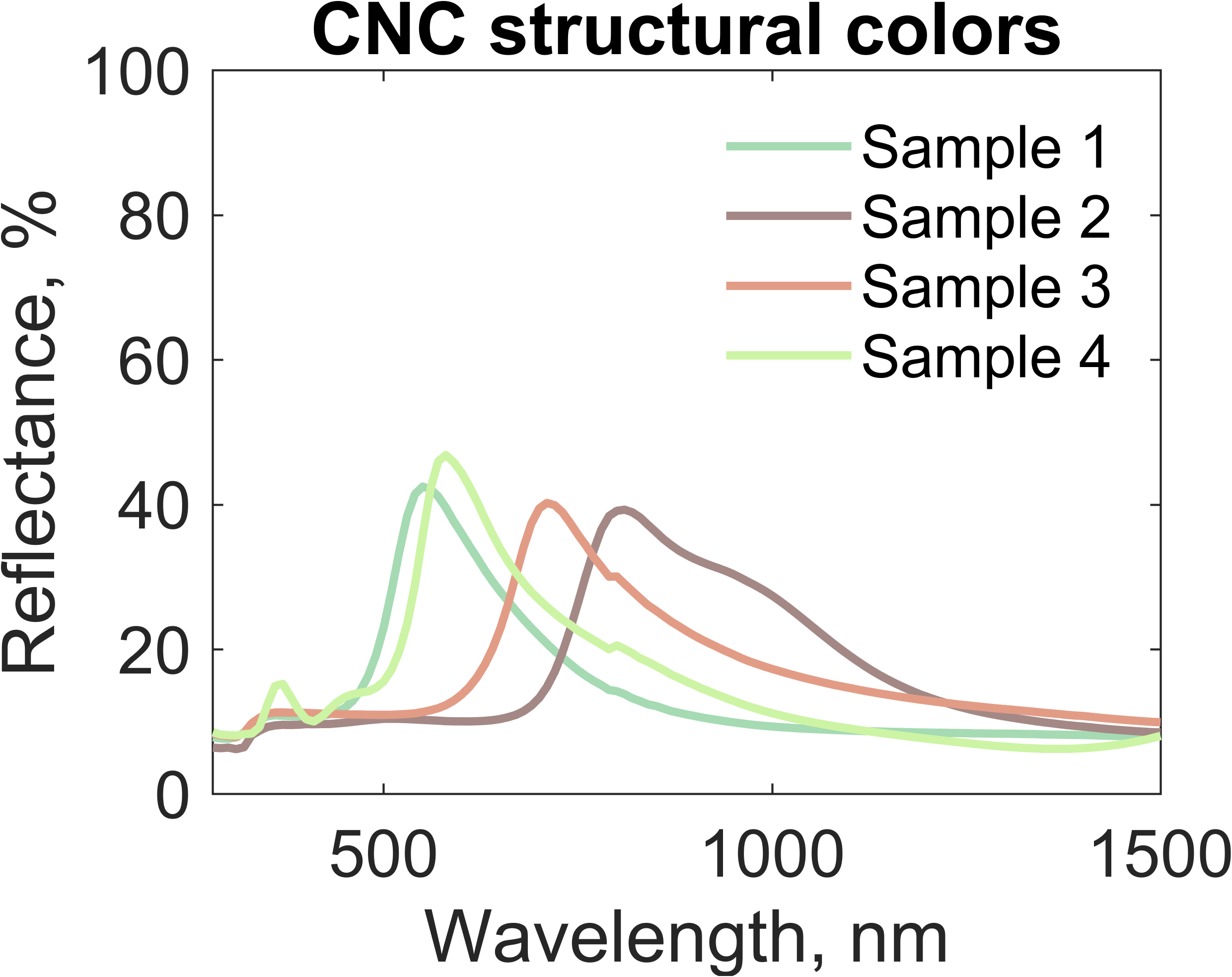}
    \caption{Measured total transmittance and total reflectance spectra of optical coatings fabricated using self-arranging CNCs.}
    \label{fig:CNC_spectra}
\end{figure}

Next, the structural colors produced using cellulose nanocrystals (CNC) were analyzed using spectrophotometry and colorimetric photography. The transmittance and reflectance spectra, shown in Fig.\,\ref{fig:CNC_spectra}, exhibit similar characteristics to the fabricated interference coatings; wide reflection peaks (150 - 250 nm width) produce bright colors without significantly lowering the transmittance. The center wavelengths correspond to the reflected colors: 550 nm for Sample 1, 810 nm for Sample 2, 710 for Sample 3, and 580 for Sample 4. Samples 1-3 exhibit transmittance exceeding 80 \% for wavelengths besides that of the target color, while Sample 4 suffers from higher absorption losses (e.g. at 1000 nm \emph{A} = 1 - \emph{R} - \emph{T} $>$ 20  \%).

\begin{figure}[h!]
    \centering
    \includegraphics[width=1\columnwidth]{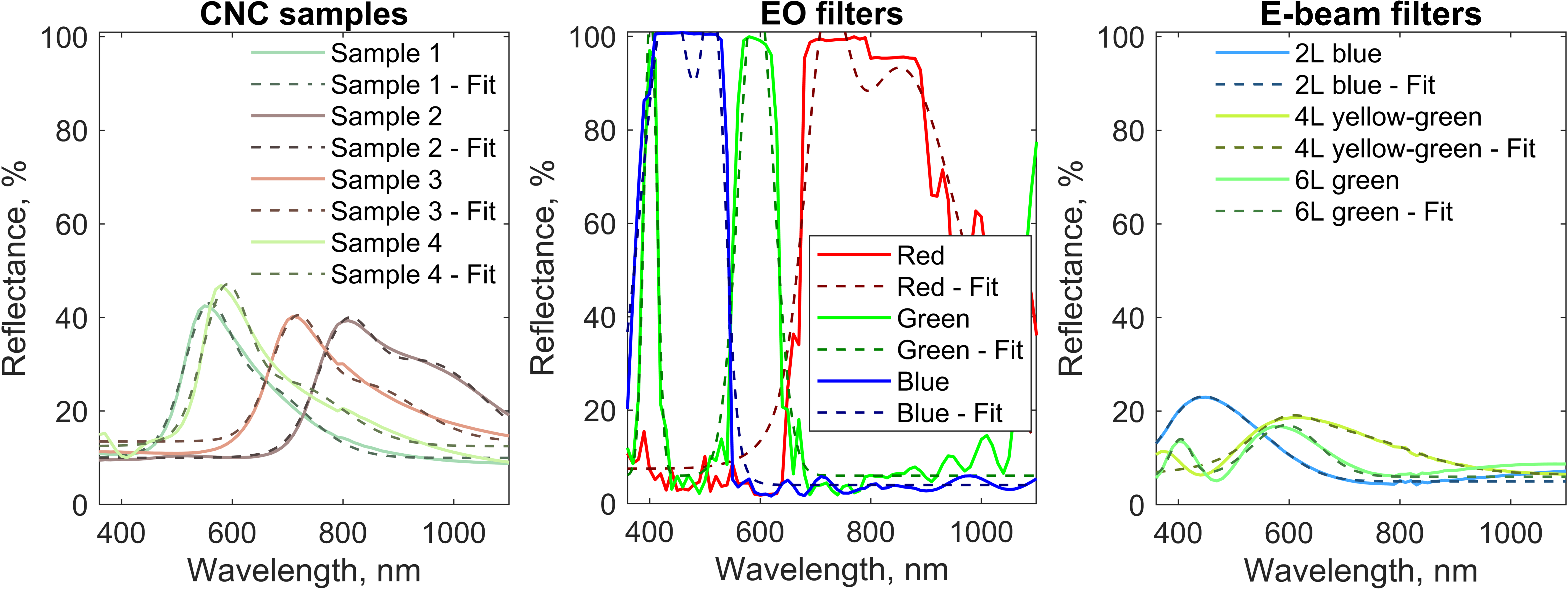}
    \caption{Total reflectance spectra of CNC samples, commercial (EO) filters, and E-beam coatings fitted with two Gaussian functions.}
    \label{fig:Reflectance_Fitted}
\end{figure}

Next, the reflectance spectra were used to extract the color generation parameters. We approximated the fitted reflectance as the sum of two Gaussian functions, as:

\begin{equation}
    \label{eq:reflectance_fit}
    R_{fit}(\lambda) = b + k_1 e^{-(\frac{\lambda-\lambda_{c,1}}{\Delta\lambda_1})^2} + k_2 e^{-(\frac{\lambda-\lambda_{c,2}}{\Delta\lambda_2})^2}
\end{equation}

where $\lambda$ is the wavelength, $b$ is the baseline, $k_1$ and $k_2$ are the peak heights, $\lambda_{c,1}$ and $\lambda_{c,2}$ are the center wavelengths, and $\Delta\lambda_{1}$ and $\Delta\lambda_{2}$ are the peak widths, often referred to as the full width at half maximum (FWHM). The results of the fitting are shown in Fig.\,\ref{fig:Reflectance_Fitted}. 

The fitting parameters, shown in Table \ref{tab:reflectance_fitted_parameters}, indicated certain trends among the three filter types. The CNC coatings demonstrated relatively high baseline reflectance (10 - 13 \%) with moderate peak heights (12 - 27 \%) and broad peak widths ($\Delta\lambda_{1}$: 56 - 68 nm, $\Delta\lambda_{2}$: 126 - 160 nm). This enables the CNCs to produce bright colors for certain viewing angles (specular reflection) without causing excessive energy losses. The commercial reflective color coatings from EO, however, demonstrated low baseline values (4 - 7 \%) with much higher peak heights (60 - 100 \%) and narrow peak widths ($\Delta\lambda_{1}$: 44 - 70 nm, $\Delta\lambda_{2}$: 16 - 28) with the one exception being the red EO coating, which had $\Delta\lambda_{2}$ = 153 nm. This coloration method, which uses narrow peak widths and high peak heights, results in high color brightness as well as high reflected energy. Narrow peaks with low baseline reflectance can generate bright colors if the peaks are sufficiently high, and they are also the theoretically most energy-efficient way for producing a given color lightness and chroma \cite{halme2019theoretical}. The interference coatings fabricated using electron beam deposition demonstrated low baseline values ($\approx$6\%) with low peak heights (5 - 13 \%). Despite the relatively broad peak widths, the fabricated coatings provided darker colors, as a result of the low overall amount of reflected VIS light.

\begin{table}[h]
    \centering
    \begin{adjustbox}{width=1\textwidth}
    \begin{tabular}{|l|c|c|c|c|c|c|c|}
      \hline
        Sample & $b$, \% & $k_1$, \% & $\lambda_{c,1}$, nm & $\Delta\lambda_{1}$, nm & $k_2$, \% & $\lambda_{c,2}$, nm & $\Delta\lambda_{2}$, nm  \\  \hline
        CNC Sample 1 & 10.0 & 22.4 & 553.0 & 56.1 & 16.3 & 638.9 & 126.6 \\ 
        CNC Sample 2 & 10.0 & 20.1 & 801.0 & 67.9 & 20.6 & 943.6 & 160.0 \\ 
        CNC Sample 3 & 13.5 & 21.2 & 712.9 & 63.3 & 12.1 & 833.8 & 137.3 \\ 
        CNC Sample 4 & 12.5 & 27.1 & 586.7 & 56.1 & 13.6 & 696.1 & 139.3 \\ 
        EO Red Filter & 7.5 & 59.9 & 717.7 & 51.9 & 85.8 & 852.7 & 152.8 \\ 
        EO Green Filter & 6.0 & 100.0 & 593.0 & 44.1 & 99.7 & 401.7 & 16.2 \\ 
        EO Blue Filter & 4.0 & 99.6 & 436.5 & 70.8 & 76.0 & 517.2 & 28.6 \\ 
        2-layer blue E-beam coating & 5.0 & 12.9 & 422.8 & 81.3 & 10.3 & 519.2 & 104.1 \\ 
        4-layer yellow-green E-beam coating & 6.5 & 5.0 & 590.2 & 75.7 & 9.1 & 677.2 & 187.1 \\ 
        6-layer green E-beam coating & 6.0 & 11.0 & 594.5 & 84.8 & 8.0 & 404.5 & 27.2 \\       \hline
    \end{tabular}
    \end{adjustbox}
    \caption{The color generation parameters, expressed by fitting the reflectance spectra of the coatings by two gaussian functions.}
    \label{tab:reflectance_fitted_parameters}
\end{table}

\begin{figure}[h!]
    \centering
    \includegraphics[width=0.8\columnwidth]{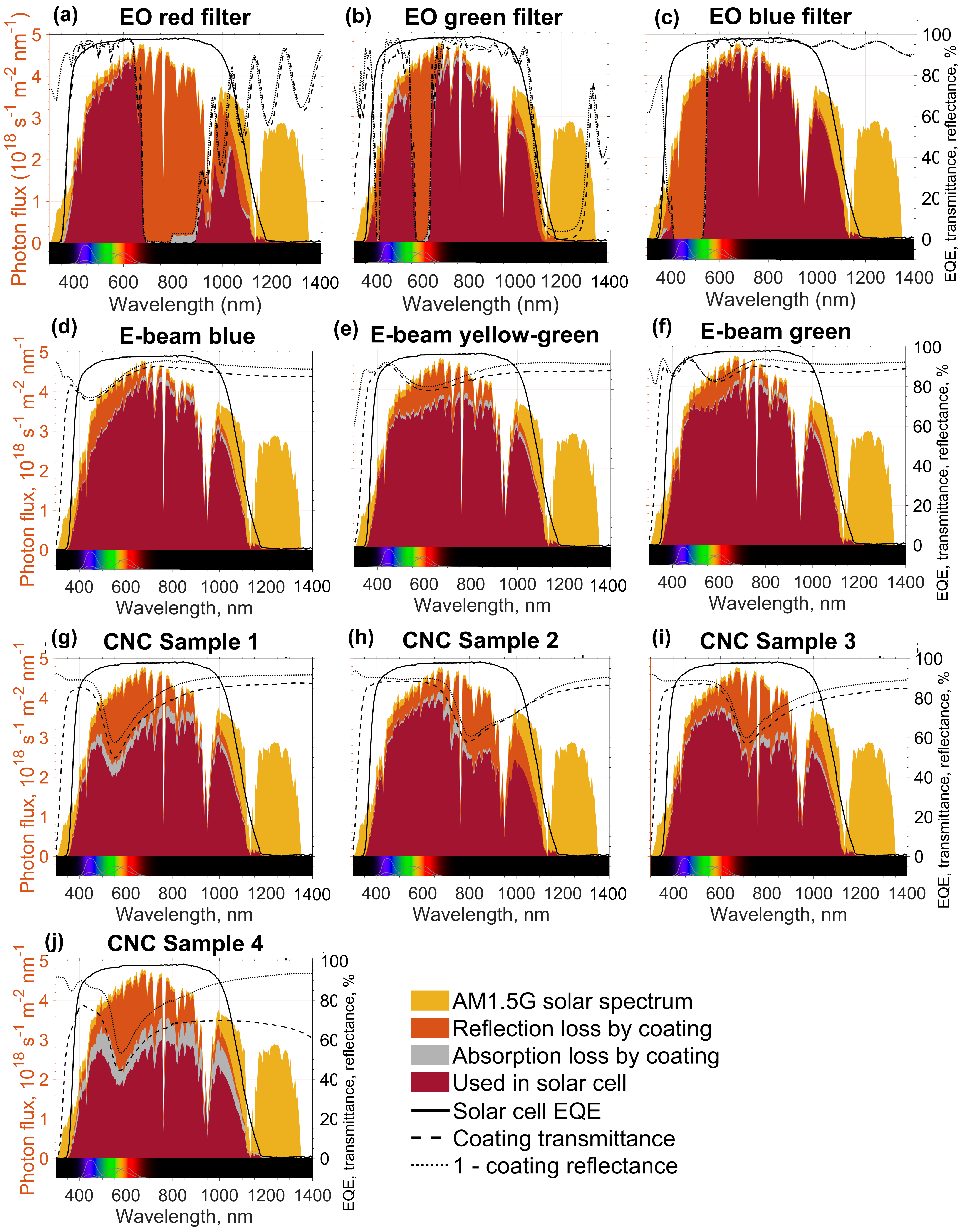}
    \caption{Energy losses due to the studied optical coatings calculated based on the total reflectance and total transmittance spectra in combination with the measured spectral response of the silicon SJSC. The following coatings were studied: (a) EO red filter, (b) EO green filter, (c) EO blue filter, (d) 2-layer blue E-beam coating, (e) 4-layer yellow-green E-beam coating, (f) 6-layer green E-beam coating, (g) CNC Sample 1, (h) CNC Sample 2, (i) CNC Sample 3, and (j) CNC Sample 4.}
    \label{fig:Loss_Figure}
\end{figure}

An analysis of the transmittance and reflectance spectra was performed using the measured external quantum efficiency (EQE) of the SJSC to identify the different energy losses in the optical coating. The losses can be divided into reflection losses (color losses) and absorption losses. The detailed spectra demonstrating the losses are displayed in Fig.\,\ref{fig:Loss_Figure} for the structural color samples, which includes the CNC coatings, commercial EO filters, and the E-beam coatings. The E-beam coatings filters provided the highest transmitted energy to the solar cell, while the commercial EO filters resulted in the largest reflection losses. Both the E-beam coatings and commercial EO filters exhibited near-zero absorption losses. The CNC coatings showed moderate reflection losses and low absorption losses, while still transmitting sunlight sufficiently in the wavelength ranges that can be absorbed by the solar cell. A similar study for the MJSC was not feasible because its EQE spectrum was not known and could not be measured by the EQE system used in this work.

The optical measurements were performed at normal incidence and light was collected using an integrating sphere. As color coatings are intended for use outdoors, where sunlight is incident at varying angles, further studies using goniometer-based spectral transmittance measurements are necessary to evaluate the average annual energy generation.

\subsection{Color characterization}

\begin{figure}[H]
    \centering
    \includegraphics[width=0.5\columnwidth]{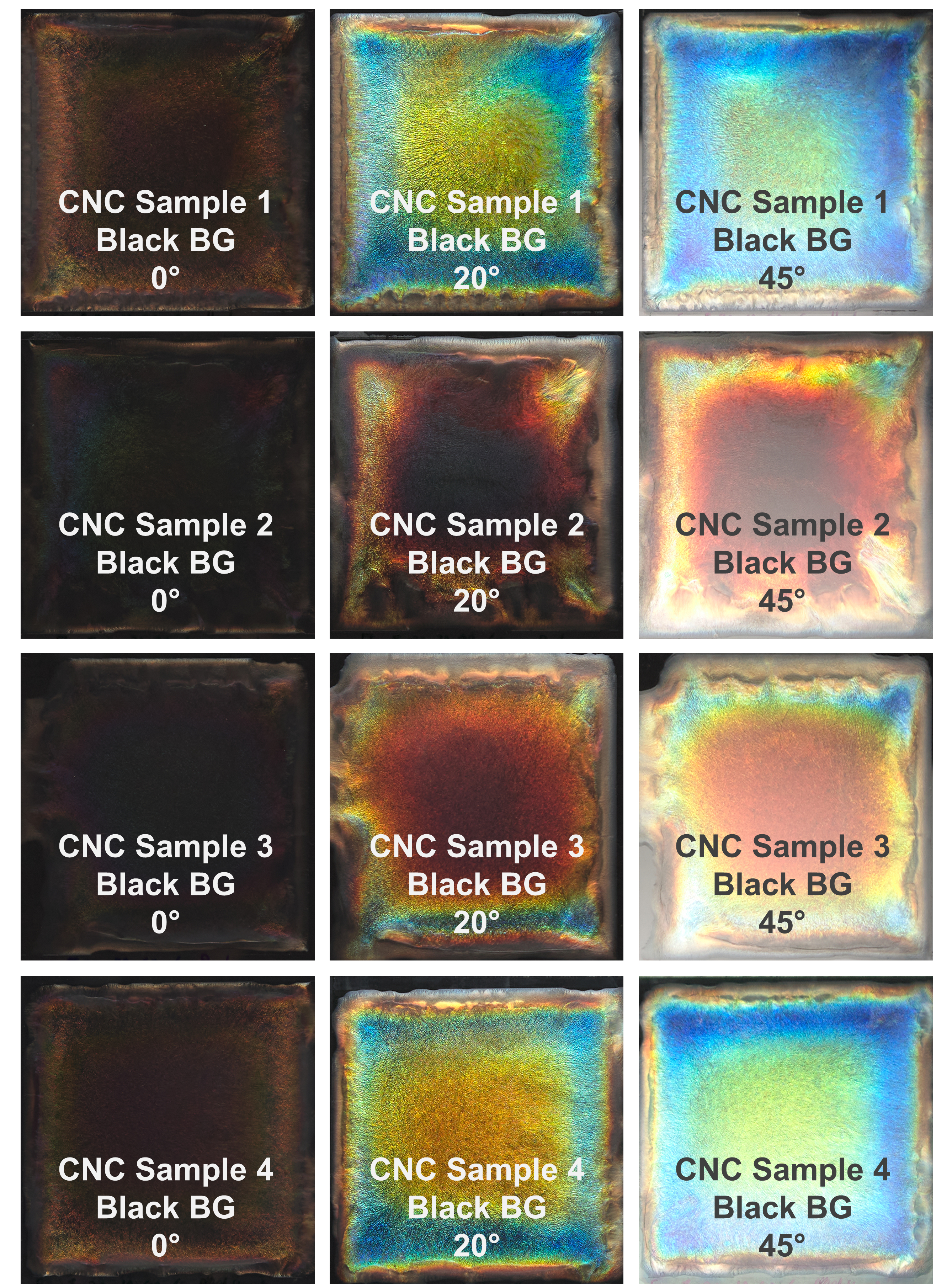}
    \caption{Photographs of the four CNC samples taken at varying angles against a diffuse black background.}
    \label{fig:CNC_Samples_Black_BG}
\end{figure}

The structural color arising from the CNCs generates a strong iridescence effect, in which the reflected color shifts depending on the viewing angle \cite{frka2019angular,schutz2020equilibrium}. The angular dependency was characterized by photographing CNC films from different angles (Fig. \ref{fig:photography_setup}) and subsequently extracting the color coordinates (Table \ref{tab:CNC_Colors}). The photography was performed with diffuse black backgrounds (Fig. \ref{fig:CNC_Samples_Black_BG}). The photography was also performed with diffuse white backgrounds, resulting in the coordinates in Table \ref{tab:CNC_Colors}. The coordinates are reported in the $L^*a^*b^*$ color space, where the $L^*$ parameter corresponds to the perceptual color lightness.

Previous work \cite{schutz2020equilibrium} show that the reflected colour is blue-shifted as the viewing angle increases. However, this only applies for specular reflections, where the illumination and viewing angles are equal. When moving away from specular reflections, the colour is red-shifted for both increasing and decreasing observation angle, as thoroughly explained by Frka-Petesic et al.\cite{frka2019angular}. This direction independence also applies to a loss of brightness when the angle increases. Figure 4 clearly shows that the colour is increasingly red-shifted when the viewing direction moves away from the direction of the specular reflection, and the $L^*$ color coordinate in Table \ref{tab:CNC_Colors} indicates that the reflected color becomes lighter when moving towards the specular reflection. This allows producing a vivid color to the observer while transmitting sufficient light to the solar cells. 
When observing the specular reflection, the color difference between the photos taken with the black and white backgrounds becomes marginal, enabling full concealment of black solar modules underneath. However, when viewed from other than the specular direction, the concealment is less effective. For example, viewing the solar cell from the direction of the surface normal (0$^\circ$) when the sunlight is incident on it at 45$^\circ$ angle, as would be approximately the case when viewing a South-facing vertical PV facade at noon in the Northern Europe, the CNC coatings would reveal the color of the underlying black solar cells.

Due to the structural origin of the color reflected by CNC coatings, the color is retained even after prolonged exposure to sunlight. This was shown experimentally by Klockars et al. \cite{klockars2019asymmetrical} where both CNC films and dyed reference textiles were exposed to six months worth of sunlight. This provides another major advantage over absorption-based methods to produce color, where photodegradation quickly leads to fading of the color.

\begin{table}[h]
    \centering
    \begin{adjustbox}{width=0.85\textwidth}
    \begin{tabular}{|l|c|c|c|c|c|c|c|c|}
    \hline
         &  \multicolumn{4}{c}{Diffuse black background}  &  \multicolumn{4}{c}{Diffuse white background} \\
         \hline
      Light: 45$^\circ$, observer: 0$^\circ$ & $L*$ & $a*$ & $b*$ & Color & $L*$ & $a*$ & $b*$ & Color  \\
      \hline
    CNC Sample 1 & 13.7 & 5.7 & 3.7 &  \cellcolor{ST_1_0deg_B} & 44.2 & 11.0 & -17.6 &  \cellcolor{ST_1_0deg_W} \\
    CNC Sample 2 & 11.8 & 0.0 & 2.2 & \cellcolor{ST_2_0deg_B}   & 55.1 & 3.3 & 0.6 &  \cellcolor{ST_2_0deg_W} \\
    CNC Sample 3 & 12.7 & 0.0 & 0.0 &  \cellcolor{ST_3_0deg_B}  & 54.4 & -3.6 & -0.7 &  \cellcolor{ST_3_0deg_W} \\ 
    CNC Sample 4 & 14.2 & 6.1 & 1.7 & \cellcolor{ST_4_0deg_B}   & 42.1 & 3.6 & -16.9 &  \cellcolor{ST_4_0deg_W} \\ 
    \hline
    Light: 45$^\circ$, observer: 20$^\circ$ & $L*$ & $a*$ & $b*$ & Color & $L*$ & $a*$ & $b*$ & Color  \\
    \hline
    CNC Sample 1 & 60.5 & -12.4 & 52.6 &  \cellcolor{ST_1_20deg_B}  & 68.0 & -5.5 & 23.4 &  \cellcolor{ST_1_20deg_W}  \\ 
    CNC Sample 2 & 17.6 & 2.4 & -1.3 &   \cellcolor{ST_2_20deg_B} & 50.6 & 2.4 & -0.9 &  \cellcolor{ST_2_20deg_W}  \\ 
    CNC Sample 3 & 23.8 & 19.8 & 8.4 &  \cellcolor{ST_3_20deg_B}  & 51.0 & 2.6 & 1.6 &  \cellcolor{ST_3_20deg_W}  \\ 
    CNC Sample 4 & 55.5 & 13.7 & 56.2 &  \cellcolor{ST_4_20deg_B}  & 63.7 & 11.8 & 26.2 &  \cellcolor{ST_4_20deg_W}  \\ 
    \hline
    Light: 45$^\circ$, observer: 45$^\circ$ & $L*$ & $a*$ & $b*$ & Color & $L*$ & $a*$ & $b*$ & Color  \\
    \hline
    CNC Sample 1 & 82.6 & -23.6 & 13.5 & \cellcolor{ST_1_45deg_B}   & 93.6 & -12.9 & 3.6 &  \cellcolor{ST_1_45deg_W}  \\ 
    CNC Sample 2 & 59.2 & 10.3 & 5.1 & \cellcolor{ST_2_45deg_B}   & 74.0 & 9.1 & 4.5 &  \cellcolor{ST_2_45deg_W}  \\ 
    CNC Sample 3 & 71.0 & 25.0 & 22.6 & \cellcolor{ST_3_45deg_B}   & 75.7 & 18.2 & 14.4 &  \cellcolor{ST_3_45deg_W}  \\ 
    CNC Sample 4 & 91.9 & -23.5 & 34.2 & \cellcolor{ST_4_45deg_B}   & 94.0 & -19.1 & 24.6 &  \cellcolor{ST_4_45deg_W}  \\ 
    \hline
    \end{tabular}
    \end{adjustbox}
    \caption{Names and $L^*a^*b^*$ color coordinates of the absorption based optical filters used in the study. The color coordinates are obtained by color-corrected photography against a diffuse white background.}
    \label{tab:CNC_Colors}
\end{table}

\subsection{Electrical characterization}

\begin{figure}[h!]
    \centering
    \includegraphics[width=0.7\columnwidth]{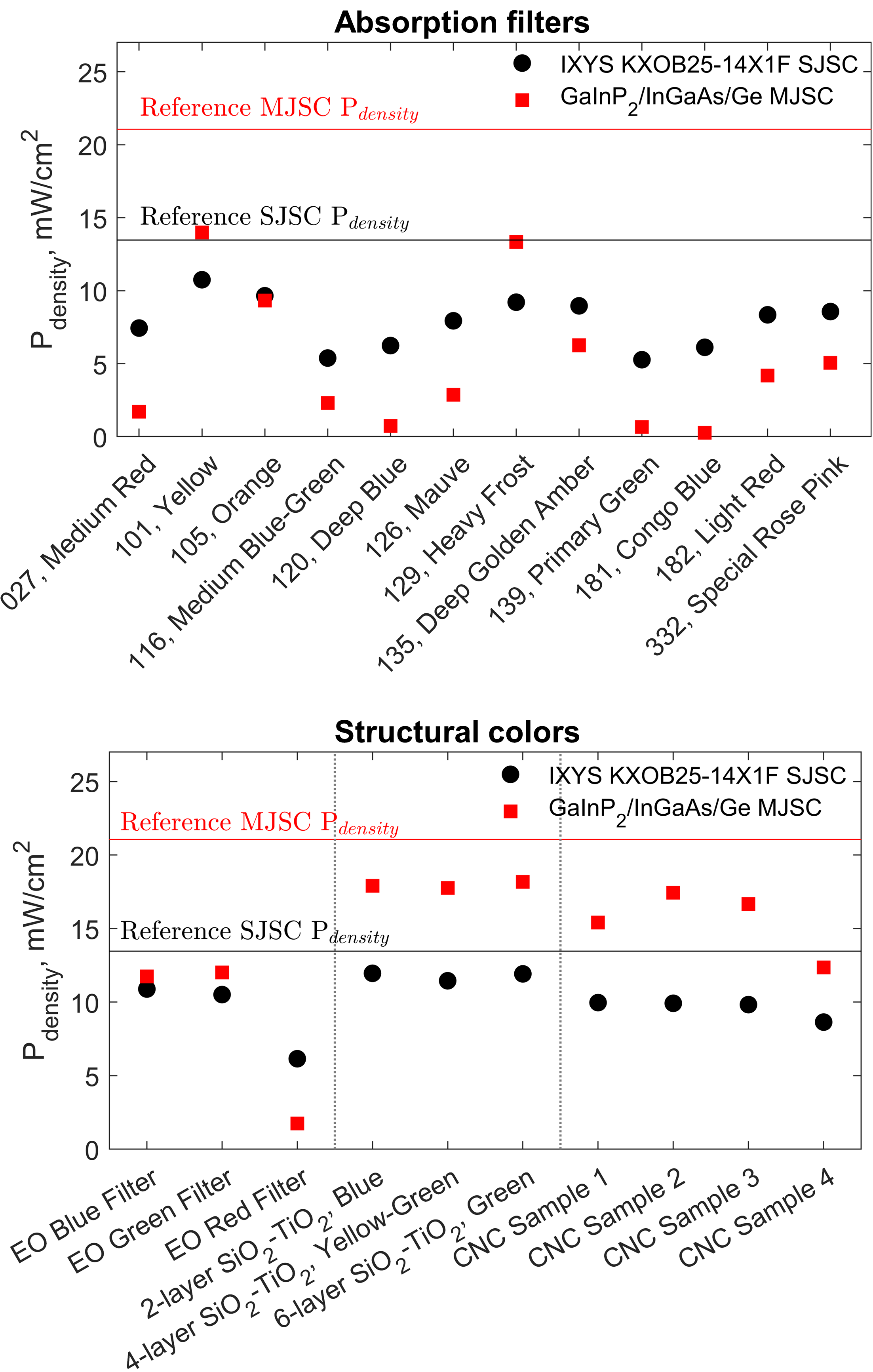}
    \caption{The power densities (P$_{density}$, mW/cm$^2$) of the SJSC and MJSC covered with absorption filters (top) and structural color coatings (bottom) relative to the reference values of the bare modules.}
    \label{fig:Pdensity}
\end{figure}

Following the optical characterization, the absorption filters and structural color coatings were used in combination with a SJSC and a MJSC, and the current-voltage (IV) characteristics were measured under 1 Sun illumination (AM1.5G spectrum, 1000 W/m$^2$ irradiance). The solar cell parameters for the SJSC and MJSC covered by the various optical coatings are reported in Tables \ref{tab:IV_Data_SJSC} and \ref{tab:IV_Data_MJSC}, and the power densities relative to the reference solar cells are shown in Fig.\,\ref{fig:Pdensity}. The coatings can be compared based on the solar cell short-circuit current ($J_{SC}$), open-circuit voltage ($V_{OC}$), fill factor ($FF$), and power density ($P_{density}$). Another metric to compare the SJSC and MJSC is the power density of the cell covered with an absorption filter relative to that of the reference cell, defined as:

\begin{equation}
    P_{relative} = \frac{P_{coating}}{P_{reference}}  .
\end{equation}

\begin{table}[h!]
    \centering
    \begin{adjustbox}{width=1\textwidth}
    \begin{tabular}{|l|c|c|c|c|c|}
      \hline
        Sample & J$_{SC}$ (mA/cm$^2$) & V$_{OC}$ (V) & FF & P$_{density}$ (mW/cm$^2$) & P$_{relative}$ \\       \hline
        Reference & 28.57 & 0.67 & 0.58 & 13.47 & 1.00 \\       \hline
        LF 027, Medium Red & 16.86 & 0.63 & 0.70 & 7.44 & 0.55 \\ 
        LF 101, Yellow & 23.54 & 0.65 & 0.70 & 10.75 & 0.80 \\ 
        LF 105, Orange & 21.37 & 0.65 & 0.70 & 9.65 & 0.72 \\ 
        LF 116, Medium Blue-Green & 12.67 & 0.61 & 0.69 & 5.38 & 0.40 \\ 
        LF 120, Deep Blue & 14.45 & 0.62 & 0.70 & 6.24 & 0.46 \\ 
        LF 126, Mauve & 17.96 & 0.63 & 0.70 & 7.94 & 0.59 \\ 
        LF 129, Heavy Frost & 20.57 & 0.64 & 0.70 & 9.21 & 0.68 \\ 
        LF 135, Deep Golden Amber & 20.04 & 0.64 & 0.70 & 8.96 & 0.67 \\ 
        LF 139, Primary Green & 12.47 & 0.61 & 0.69 & 5.27 & 0.39 \\ 
        LF 181, Congo Blue & 14.23 & 0.62 & 0.69 & 6.12 & 0.45 \\ 
        LF 182, Light Red & 18.81 & 0.63 & 0.70 & 8.35 & 0.62 \\ 
        LF 332, Special Rose Pink & 19.27 & 0.63 & 0.70 & 8.57 & 0.64 \\       \hline
        EO Reflective Filter Red & 13.72 & 0.64 & 0.70 & 6.15 & 0.46 \\ 
        EO Reflective Filter Green & 22.26 & 0.66 & 0.71 & 10.51 & 0.78 \\ 
        EO Reflective Filter Blue & 22.93 & 0.67 & 0.71 & 10.89 & 0.81 \\       \hline
        2-layer SiO$_2$-TiO$_2$ Blue & 25.05 & 0.67 & 0.71 & 11.96 & 0.89 \\ 
        4-layer SiO$_2$-TiO$_2$ Yellow-Green & 24.21 & 0.66 & 0.71 & 11.45 & 0.85 \\ 
        6-layer SiO$_2$-TiO$_2$ Green & 25.13 & 0.67 & 0.71 & 11.92 & 0.89 \\       \hline
        CNC Sample 1 & 21.54 & 0.66 & 0.71 & 9.97 & 0.74 \\ 
        CNC Sample 2 & 21.54 & 0.65 & 0.71 & 9.92 & 0.74 \\ 
        CNC Sample 3 & 21.39 & 0.65 & 0.70 & 9.83 & 0.73 \\ 
        CNC Sample 4 & 18.72 & 0.65 & 0.70 & 8.64 & 0.64 \\ 
    \hline
    \end{tabular}
    \end{adjustbox}
    \caption{The SJSC parameters at 1 Sun illumination when covered with the various optical coatings.}
    \label{tab:IV_Data_SJSC}
\end{table}

\begin{table}[h!]
    \centering
    \begin{adjustbox}{width=1\textwidth}
    \begin{tabular}{|l|c|c|c|c|c|}
      \hline
        Sample & J$_{SC}$ (mA/cm$^2$) & V$_{OC}$ (V) & FF & P$_{density}$ (mW/cm$^2$) & P$_{relative}$ \\       \hline
        Reference & 12.42 & 2.46 & 0.72 & 21.06 & 1.00 \\      \hline
        LF 027, Medium Red & 2.53 & 2.22 & 0.30 & 1.70 & 0.08 \\ 
        LF 101, Yellow & 8.50 & 2.37 & 0.70 & 14.00 & 0.66 \\ 
        LF 105, Orange & 6.31 & 2.35 & 0.63 & 9.32 & 0.44 \\ 
        LF 116, Medium Blue-Green & 2.72 & 2.22 & 0.38 & 2.30 & 0.11 \\ 
        LF 120, Deep Blue & 1.85 & 2.05 & 0.19 & 0.74 & 0.04 \\ 
        LF 126, Mauve & 3.21 & 2.27 & 0.39 & 2.87 & 0.14 \\ 
        LF 129, Heavy Frost & 8.12 & 2.36 & 0.70 & 13.34 & 0.63 \\ 
        LF 135, Deep Golden Amber & 4.88 & 2.34 & 0.55 & 6.26 & 0.30 \\ 
        LF 139, Primary Green & 1.76 & 1.99 & 0.19 & 0.66 & 0.03 \\ 
        LF 181, Congo Blue & 1.40 & 1.40 & 0.14 & 0.27 & 0.01 \\ 
        LF 182, Light Red & 3.90 & 2.31 & 0.46 & 4.18 & 0.20 \\ 
        LF 332, Special Rose Pink & 4.37 & 2.34 & 0.50 & 5.06 & 0.24 \\      \hline
        EO Reflective Filter Red & 1.63 & 2.29 & 0.47 & 1.75 & 0.08 \\ 
        EO Reflective Filter Green & 7.31 & 2.42 & 0.68 & 12.02 & 0.57 \\ 
        EO Reflective Filter Blue & 7.55 & 2.42 & 0.64 & 11.75 & 0.56 \\      \hline
        2-layer SiO$_2$-TiO$_2$ Blue & 11.08 & 2.42 & 0.67 & 17.91 & 0.85 \\ 
        4-layer SiO$_2$-TiO$_2$ Yellow-Green & 10.97 & 2.41 & 0.67 & 17.76 & 0.84 \\ 
        6-layer SiO$_2$-TiO$_2$ Green & 11.34 & 2.40 & 0.67 & 18.17 & 0.86 \\      \hline
        CNC Sample 1 & 9.03 & 2.39 & 0.72 & 15.42 & 0.73 \\ 
        CNC Sample 2 & 10.03 & 2.39 & 0.73 & 17.43 & 0.83 \\ 
        CNC Sample 3 & 9.47 & 2.39 & 0.74 & 16.68 & 0.79 \\ 
        CNC Sample 4 & 8.17 & 2.39 & 0.63 & 12.36 & 0.59 \\ 
    \hline
    \end{tabular}
    \end{adjustbox}
    \caption{The MJSC parameters at 1 sun illumination when covered with the various optical coatings.}
    \label{tab:IV_Data_MJSC}
\end{table}

The SJ crystalline silicon solar cell had an initial power density of 13.47 mW/cm$^2$ while the triple junction GaInP$_2$/InGaAs/Ge solar cell had an initial power density of 21.06 mW/cm$^2$ at an irradiated power of approximately 100 mW/cm$^2$, which are typical values for commercial solar cells. The efficiency of the MJSC significantly exceeds that of the SJSC while coming at significantly higher costs as well as having restrictions on maintaining current matching between the individual sub-cells.

When the electrical characterization was performed with the absorption filters covering the solar modules, the remaining power density values were vastly different. The SJSC outperformed the MJSC in almost all colors when used in combination with absorption color filters, as shown in Fig.\,\ref{fig:Pdensity} (top). This results from the strong current mismatch due to significantly lower transmittance at all wavelengths below 500 - 600 nm, and therefore weaker absorption in the sub-cell encompassing the UV-VIS region. The V$_{OC}$ of the SJSC remains relatively unchanged for all colors (0.61 - 0.65 V) relative to the reference (0.67 V) while that of the MJSC varied drastically (1.4 - 2.36 V) relative to the reference (2.46 V). As with the V$_{OC}$, the $P_{relative}$ shows notably higher deterioration for the MJSC than for the SJSC due to current mismatch losses.

On the other hand, the solar cells showed greater efficiencies when covered with the structural colored samples. First, the commercial EO reflective filters resulted in P$_{relative}$ reduction by 54 \% red, 22 \% green, and 19 \% blue for the SJSC. The P$_{relative}$ reduction was significantly higher for MJSCs due to the current mismatch, which reached 92 \% for red, 43 \% for green, and 44 \% for blue. Next, the fabricated E-beam coatings were studied using the same method, exhibiting remarkably greater power densities for both the SJSC and MJSC. The P$_{relative}$ reduction ranges from 11 - 15 \% for the SJSC and 14 - 16 \% for the MJSC. Since both modules retain most of their performance when covered by the E-beam coatings, the MJSC outperforms the SJSC by producing approximately 18 mW/cm$^2$ as opposed to 12mW/cm$^2$ for the SJSC. Further optimization of the E-beam coatings can be performed to minimize the current matching losses and achieve even higher performance. Finally, the solar modules were measured with the CNC color coatings, showing highly promising results for the SJSC and MJSC. The relative power density remains high for most samples, with samples 2 and 3 (reddish brown colors) providing the most promising results. The P$_{relative}$ reduction ranges between 26 - 36 \% for the SJSC and between 17 - 41 \% for the MJSC.

Both the fabricated CNC and E-beam multilayer samples outperformed the rest due to reflecting less light, thereby producing darker reflected colors. Their coloration method utilizes broadband peaks with low peak height. On the other hand, the coloration method based on one or multiple high reflection peaks reflects brighter colors while resulting in higher losses, particularly due to exacerbating the current mismatch of the MJSC.

\section{Conclusion}

This study aimed to identify the features of efficient color coatings for colored solar cells. The features were investigated by performing a systematic experimental study of the color coatings made using different coloration methods: absorption filters, interference filters, and photonic crystals. Both experimentally fabricated and commercially-obtained optical filters were used in the study. Due to the different color and coloration methods, it would be difficult to predict the actual optical efficiency in a real application situation, where also cost and durability have to be optimized. Nevertheless, general trends were observed regarding the performance of the studied coloration methods. Moreover, the impact of the different coloration methods on the performance of single junction and tandem solar cells was addressed.

The samples were optically characterized using spectrophotometry and their energy loss components were identified. The color generation was studied by examining the reflectance parameters and by performing colorimetric photography on select samples. Structural colors were more efficient than absorption filters in reflecting colors while transmitting the remaining light. Furthermore, structural colors are suitable for BIPV as their colors do not fade over time as long as the structure is not compromised. Coatings that reflect high, narrow peaks were able to produce bright colors, at the cost of lower transmitted energy. Coatings that reflected lower but broader peaks produced darker colors, while allowing significantly more light to be transmitted to the solar cells. CNC coatings provided a suitable balance between color brightness and light transmission by reflecting wide, moderately-high peaks.

The coatings were further characterized by measuring the electrical parameters of a SJSC and a MJSC covered by the samples. The performance of the SJSC remained acceptable using the various color coatings, providing a power density of 5 - 12 mW/cm$^2$ for 100 mW/cm$^2$ illumination power. The MJSC, however, provided wide variations in power density (1 - 18 mW/cm$^2$ for the same illumination) depending on the reflected peaks' center wavelengths, heights, and widths.

Both solar modules can be successfully used towards colored solar cells if certain conditions are met. The main limitation of SJSCs is their limited spectral response, which is addressed by ensuring that the coating transmits sufficient light in the solar cell absorption region. As for MJSCs, the wider spectral response comes at the expense of current mismatch losses that significantly deteriorate the cell's performance. Accordingly, designing color coatings for MJSCs requires accounting for the current mismatch by examining the EQE of individual subcells and ensuring current matching conditions. Further examination of the current mismatch losses is necessary by characterizing the spectral response of MJSCs as well as simultaneously optimizing the color parameters and solar cell bandgap(s).

By studying the factors influencing the performance of colored solar cells, opportunities emerge to design and identify more efficient and scalable color coatings for use in BIPV. Structural colors are often used in commercial colored BIPV modules due to their low absorption losses. However, colored BIPV modules nowadays predominantly use silicon solar cells. Combining color coatings and tandem solar cells is a promising approach for high efficiency BIPV modules, particularly as novel tandem devices become cost-effective and prevalent in all PV applications.

Future studies on the correlation between experimentally measured color brightness and the impact on solar cell efficiency are necessary to optimize color coatings for both high brightness and performance. Characterization using angle-dependent photography is a useful tool to study color coatings from different viewing angles. A study on the solar cell performance for varying incidence angles is also crucial to predict the annual energy generation of colored solar cells, which could be the topic of a future work. Furthermore, the practical relevance of various color coatings should be studied for integration on BIPV facades by examining their aesthetic appeal to architects and building users.

\section{Acknowledgements}
The authors thank Jaakko Eskola for assisting in the color characterization using colorimetric photography.

\section{Funding}
This research did not receive any specific grant from funding agencies in the public, commercial, or not-for-profit sectors.

 \bibliographystyle{elsarticle-num} 
 \bibliography{references}

\end{document}